\documentclass{article}
\usepackage{spconf,amsmath,graphicx}
\usepackage{multirow}
\usepackage{comment}
\usepackage{amsmath}
\usepackage{amssymb}
\usepackage{amsfonts}
\usepackage{enumerate}
\usepackage{siunitx}
\usepackage{booktabs}
\usepackage{color}
\usepackage{hyperref}
\usepackage{setspace}

\newlength\savedwidth
\newcommand{\wcline}[1]{\noalign{\global\savedwidth\arrayrulewidth\global\arrayrulewidth 1.0pt} \cline{#1}
\noalign{\global\arrayrulewidth\savedwidth}}
\usepackage{bm}
\newcommand{\bmit}[1]{{\mbox{\boldmath $#1$}}}

\newcommand{\norm}[1]{\left\lVert#1\right\rVert}
\newcommand{\trans}[1]{#1^\mathsf{T}}

\title{Environmental sound extraction using onomatopoeic words}
\name{Yuki Okamoto$^{1}$\sthanks{This work has been done during internship at Hitachi, Ltd.}\quad
       Shota Horiguchi$^{2}$\quad 
       Masaaki Yamamoto$^{2}$\quad 
       Keisuke Imoto$^{3}$\quad
       Yohei Kawaguchi$^{2}$
       }
 \address{$^1$ Ritsumeikan University, Japan \\$^2$ Hitachi, Ltd., Japan\\$^3$Doshisha University, Japan
  }

\begin{document}
\ninept
\maketitle
%
\begin{abstract}
An onomatopoeic word, which is a character sequence that phonetically imitates a sound, is effective in expressing characteristics of sound such as duration, pitch, and timbre.
We propose an environmental-sound-extraction method using onomatopoeic words to specify the target sound to be extracted.
By this method, we estimate a time-frequency mask from an input mixture spectrogram and an onomatopoeic word using a U-Net architecture, then extract the corresponding target sound by masking the spectrogram.
Experimental results indicate that the proposed method can extract only the target sound corresponding to the onomatopoeic word and performs better than conventional methods that use sound-event classes to specify the target sound.
\end{abstract}
\vspace{-7pt}
\section{Introduction}
\label{sec:intro}
\vspace{-5pt}
Environmental sounds are essential for expressive media content, e.g., movies, video games, and animation, to make them immersive and realistic.
One way to prepare a desired sound is to obtain it from an environmental sound database.
However, the number of databases currently available is very limited \cite{Imoto_AST_2018}, so the desired sound is not always in the database.
On the other hand, there is a large amount of unlabeled environmental sounds on the Internet, but it is not easy to expand the database because it requires rich domain knowledge and taxonomy.

Even if the database became large, its usability might decrease because it would also require users to have domain knowledge.
Intuitive methods for sound retrieval have been proposed.
For example, vocal imitation \cite{zhang_CHIIR_2020, Zhang_ICASSP_2016,Kim_ICASSP_2019, Zhang_TASLP_2019} and onomatopoeic words \cite{Ikawa_DCASE_2018} were used as search queries in some sound retrieval systems.
It has also been reported that user satisfaction is high when using an intuitive sound-retrieval technique \cite{zhang_CHIIR_2020,Zhang_ICASSP_2016}.
Therefore, it would also be useful for content creators if they can extract a desired sound intuitively.
\begin{figure}[t!]
\centering
\includegraphics[scale=0.9]{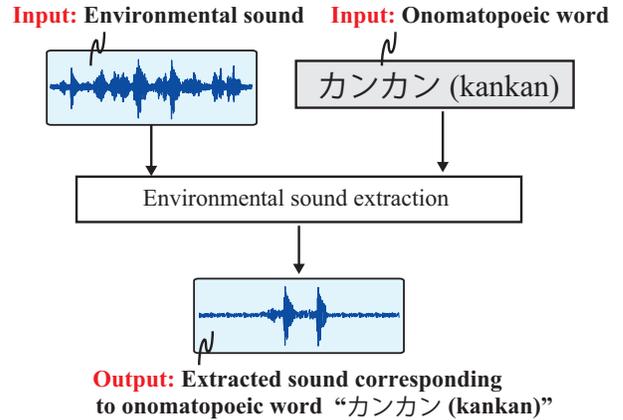}
\vspace{-2pt}
\caption{Overview of environmental sound extraction using an onomatopoeic word. The word ``kankan'' is often used in Japanese to represent hitting sounds.}
\label{fig:overview_method}
\end{figure}
\begin{figure*}[t!]
\centering
\resizebox{\linewidth}{!}{%
\includegraphics{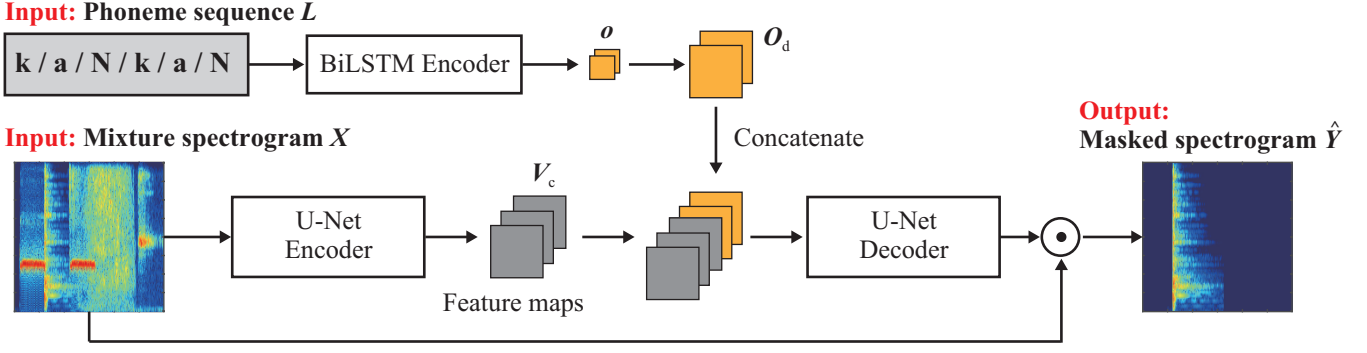}%
}
\caption{Detailed architecture of proposed environmental-sound-extraction method using an onomatopoeic word.}
\label{fig:propsed_method}
\end{figure*}

We propose an environmental-sound-extraction method using an onomatopoeic word, which is a character sequence that phonetically imitates a sound.
It has been shown that onomatopoeic words are effective in expressing the characteristics of sound \cite{Lemaitre_JASA_2018,Sundaram_AAAI_2006} such as sound duration, pitch, and timbre.
Onomatopoeic words are also advantageous in terms of low labeling cost since they do not require domain knowledge and taxonomy for labeling.
In our proposed method, therefore, uses an onomatopoeic word is used to specify the sound to extract from a mixture sound, as shown in Fig.~\ref{fig:overview_method}.
We used a U-Net architecture \cite{Ronneberger_MICCAI_2015}, which has been used in various source-separation and sound-extraction studies \cite{Mesequer_ISMIR_2019, Sudo_IROS_2019, Jansson_ISMIR_2017, Kong_ICASSP_2020}, to estimate the time-frequency mask of the target sound.
To the best of our knowledge, there has been no study on extracting only specific sound by using an onomatopoeic word.

The rest of the paper is organized as follows.
In Sec.~\ref{sec:conventional}, we describe related work on environmental sound extraction.
In Sec.~\ref{sec:propose}, we present our proposed method for extracting environmental sounds using an onomatopoeic word from an input mixture.
In Sec.~\ref{sec:experiments}, we discuss experiments we conducted on the effectiveness of our proposed method compared with baseline methods that use class labels to specify the target sound.
Finally, we summarize and conclude this paper in Sec.~\ref{sec:conclusion}.

\section{RELATED WORK}
\label{sec:conventional}
\vspace{-5pt}
Methods of environmental sound extraction and separation using deep learning have been developed \cite{Sudo_IROS_2019, ochiai_INTERSPEECH_2020, Lee_ISMIR_2019, Kavalerov_WASPPA_2019}.
Sudo et al. developed an environmental-sound-separation method based on U-Net architecture \cite{Sudo_IROS_2019}.
A similar method using U-Net was also proposed for source separation \cite{Mesequer_ISMIR_2019, Slizovskaia_ICASSP_2109}.
Ochiai et al. used Conv-TasNet \cite{Luo_TASLP_2019}, which was originally proposed for speech separation, to extract only the sounds of specific sound events \cite{ochiai_INTERSPEECH_2020}.
These methods use the sound-event class as the input to specify the desired sound.
However, environmental sounds have various characteristics that cannot be described as a sound class, such as sound duration, pitch, and timbre.
For example, if the ``whistle sound'' class is defined regardless of the pitch, it is not possible for a conventional method to extract only the sound of the desired pitch.
One possible solution is to define more fine-grained sound-event classes, e.g., ``high-pitched whistle sound'' and ``low-pitched whistle sound.''
However, this is impractical because the labeling cost will increase.
Even if we could define such fine-grained sound-event classes, there would always be intra-class variation, and we have no way to distinguish between them.
Therefore, the method of conditioning by sound-event class is not suitable to extract specific sounds. 

Another possible method similar to that discussed in this paper is singing voice extraction using humming \cite{Smaradis_WASPAA_2019}.
In this case, the target sound is always a human voice, so humming is sufficient to represent it.
However, in the case of environmental sound extraction, humming is insufficient to determine the target sound because it cannot express timbre, and some kinds of sounds cannot be represented by humming, e.g., plosive sounds.

\section{PROPOSED METHOD}
\label{sec:propose}
\vspace{-5pt}
\subsection{Overview of environmental sound extraction using an onomatopoeic word}
\label{subsec:overview}
\vspace{-5pt}
Our purpose was to reconstruct a target sound $\bmit{y}$ from a mixture sound $\bmit{x}$, where the target is specified by an onomatopoeic word $w$.
We estimate $\hat{\bmit{y}}$ from $\bmit{x}$ and $w$ using a nonlinear transformation $\mathsf{Extractor}(\cdot,\cdot)$ as follows:
\begin{align}
    \hat{{\bmit y}} = \mathsf{Extractor}({\bmit x}, w).
    \label{eq:train}
\end{align}
We explain this $\mathsf{Extractor}(\cdot,\cdot)$ in Sec.~\ref{subsec:training_method}.

%
%
\subsection{Proposed sound extraction method}
\label{subsec:training_method}
\vspace{-5pt}
Figure \ref{fig:propsed_method} shows the detailed architecture of the proposed method.
The method involves time-frequency mask estimation using U-Net and feature vector extraction from an onomatopoeic word.
We condition the output of the U-Net encoder with an onomatopoeic word to specify the target environmental sound to extract.
In previous studies, the target sound to be extracted was conditioned by sound-event class \cite{Mesequer_ISMIR_2019}, or further conditioned by the estimated interval of the target sound \cite{Sudo_IROS_2019}.
These studies have shown that conditioning on intermediate features after passing through convolutional neural network layers can be effective.
Thus, we also use conditioning on the intermediate features of the U-Net encoder.

The proposed method takes the following as inputs, as shown in Fig. ~\ref{fig:propsed_method}.
One is a $T$-length $F$-dimensional mixture spectrogram $\bm{X} \in \mathbb{R}^{F \times T}$ extracted from the input mixture sound ${\bmit x}$. 
The other is a one-hot encoded phoneme sequence ${\bmit L}=(\bmit{l}_1,\dots,\bmit{l}_{N})$ extracted from $w$.
The extracted acoustic feature $\bm{X}$ is fed to the U-Net encoder, which consists of $K$-stacked convolutional layers.
In each layer of the U-Net encoder, the time-frequency dimension decreases by half and the number of channels doubles.
As a result, $C\left(=2^K\right)$ feature maps are calculated as
\begin{align}
    \left[\bm{V}_1,\dots,\bm{V}_C\right]=\mathsf{UNetEncoder}\left(\bm{X}\right)\in\mathbb{R}^{F'\times T'\times C},
\end{align}
where $\bmit{V}_c\in\mathbb{R}^{F'\times T'} (c=1,\dots,C)$ denotes the feature map of the $c$-th channel.

At the same time, the phoneme sequence ${\bmit L}$ is fed to the bidirectional long short-term memory (BiLSTM) encoder.
As a result, a $D$-dimensional word-level embedding ${\bmit o}=\trans{\left[o_1,\dots,o_D\right]}\in\mathbb{R}^{D}$ that captures the entire onomatopoeic word is extracted as follows:
\begin{align}
    \bmit{o}=\mathsf{BiLSTMEncoder}\left(\bmit{L}\right)\in\mathbb{R}^{D}.\label{eq:word_embedding}
\end{align}
The extracted embedding $\bm{o}$ is stretched in the time and frequency directions to form $D$ feature maps $\left[\bmit{O}_1,\dots,\bmit{O}_D\right]$, where $\bmit{O}_d\in\mathbb{R}^{F'\times T'}$ for $d\in\left\{1,\dots,D\right\}$ is the matrix the whose elements are all $o_d$.

Finally, a time-frequency soft mask $M\in\left(0,1\right)^{F\times T}$ is estimated using the U-Net decoder, which consists of $K$-stacked deconvolutional layers.
The feature maps from the U-Net encoder and BiLSTM encoder are concatenated to be $C+D$ channels and fed to the U-Net decoder followed by the element-wise sigmoid function $\sigma\left(\cdot\right)$ as
\begin{align}
    \bmit{Z}&=\mathsf{UNetDecoder}\left(\left[\bmit{V}_1,\dots,\bmit{V}_C,\bmit{O}_1,\dots,\bmit{O}_D\right]\right)\in\mathbb{R}^{F\times T},\\
    \bm{M}&=\sigma\left(\bmit{Z}\right)\in\left(0,1\right)^{F\times T}.\label{eq:softmask}
\end{align}
The target signal in time-frequency domain $\hat{\bm{Y}}$ is then recovered by masking the input $\bm{Y}$ as
\begin{align}
    \hat{\bm{Y}}=\bm{M}\odot\bm{X}\in\mathbb{R}^{F\times T},
\end{align}
where $\odot$ is the Hadamard product.

During training, the loss function defined as root mean square error between $\hat{\bm{Y}}$ and target features $\bm{Y}\in\mathbb{R}^{F\times T}$, which is extracted from $\bm{x}$, is used:
\begin{align}
    L\left(\bm{Y},\bm{\hat{Y}}\right)&=\sqrt{\frac{1}{TF} \norm{\bm{Y} - \bm{\hat{Y}}}_F^2},
\end{align}
where $\norm{\cdot}_F$ is the Frobenius norm.

In the inference phase, we reconstruct an environmental sound wave from the masked acoustic features $\bm{\hat{Y}}$ using the Griffin–Lim algorithm \cite{Griffin_TASSP_1984}.

\section{EXPERIMENTS}
\label{sec:experiments}
\vspace{-5pt}
\subsection{Dataset construction}
\label{sec:dataset}
\vspace{-5pt}
To construct the datasets for this task, we used environmental sounds extracted from RealWorld Computing Partnership-Sound Scene Database (RWCP-SSD) \cite{Nakamura_AST_1999}.
Some sound events in RWCP-SSD are labeled in the ``event entry + ID'' format, e.g., \textit{whistle1} and \textit{whistle2}.
We created hierarchical sound-event classes by grouping labels with the same event entry, e.g., \textit{whistle}.
We first selected 44 sound events from RWCP-SSD, which we call subclasses, and grouped them into 16 superclasses.
The superclasses and subclasses used in this study are listed in Table \ref{table:class_list}.
We selected 16 types of sound events in superclass and 44 types of sound events in subclass from RWCP-SSD to construct the dataset.
The sounds in each subclass were divided as 7:2:1, used for training, validation, and evaluation, respectively.
The onomatopoeic words corresponding to each environmental sound were extracted from RWCP-SSD-Onomatopoeia \cite{okamoto_DCASE_2020}.
Each sound was annotated with more than 15 onomatopoeic words in RWCP-SSD-Onomatopoeia, and we used randomly selected three onomatopoeic words for each sound for our experiments.

We constructed the following three evaluation datasets using the selected sound events:
\begin{itemize}
\item {\bf Inter-superclass dataset}:
Each mixture sound in this dataset is composed of a target sound and interference sounds, the superclass of each is different from that of the target sound.
\item {\bf Intra-superclass dataset}:
Each mixture sound in this dataset is composed of a target sound and interference sounds, the superclass of each is the same as that of the target sound, but the subclass is different.
\item {\bf Intra-subclass dataset}:
Each mixture sound in this dataset is composed of a target sound and interference sounds, the subclass of each is the same as that of the target sound, but the onomatopoeic words are different.
\end{itemize}
The mixture sounds in each dataset were created by varying the signal-to-noise ratio (SNR) by $\{-10, -5, 0, 5, 10\}$ \si{\dB}.
The SNR between a target signal $\bm{s}_\text{target}$ and an interference signal $\bm{s}_\text{interference}$ is defined as
\begin{align}
    \mbox{SNR} = 10\log_{10}\left(\frac{\norm{\bm{s_\text{target}}}^2}{\norm{\bm{s}_\text{interference}}^2}\right).
\end{align}
The training and validation sets consisted of 7,563 and 2,160 mixture sounds, respectively.
Each evaluation set consisted of 1,107 mixture sounds for each SNR. 
The audio clips for these sets were randomly selected from RWCP-SSD. 
\begin{table}[t!]
\caption{Experimental conditions}
\vspace{2pt}
\label{table:experiment}
\centering
\begin{tabular}{@{}ll@{}}
    \wcline{1-2}
     &\\[-8pt]
    Mixture-sound length & \SI{5}{\second} \\
    Sampling rate & \SI{16}{kHz}\\
    Waveform encoding & 16-bit linear PCM \\
    \cline{1-2}
    &\\[-8pt]
    \# of U-Net encoder blocks& 4\\ 
    \# of U-Net decoder blocks& 4\\ 
    \# of BiLSTM encoders & 1 \\
    \# of units in BiLSTM encoder & 512\\
    Batch size & 8 \\
    Optimizer  & RAdam \cite{RAdam_ICLR_2020}\\
    \cline{1-2}
    &\\[-8pt]
    Acoustic feature & Amplitude spectrogram\\
    Window length for FFT &  \SI{128}{\ms} (2,048 samples) \\
    Window shift for FFT & \SI{32}{\ms} (512 samples) \\
    \wcline{1-2}
\end{tabular}
\end{table}
\begin{table}[t]
    \caption{Superclass and subclass sound events used in this study}
    \vspace{2pt}
    \label{table:class_list}
    \centering
    \resizebox{\linewidth}{!}{%
    \begin{tabular}{@{}ll|ll@{}}
        \wcline{1-4}
        &\\[-8pt]
        Superclass & Subclass&Superclass&Subclass\\
        \cline{1-4}
        &\\[-8pt]
        \textbf{metal}&metal05, metal10,&\textbf{bells}&bells1, bells2, bells3,\\
        &metal15&&bells4, bells5\\
        \textbf{dice}&dice1, dice2, dice3&\textbf{coin}&coin1, coin2, coin3\\
        \textbf{bottle}&bottle1, bottle2&\textbf{coins}&coins1, coins2, coins3,\\
        \textbf{cup}&cup1, cup2&&coins4, coins5\\
        \textbf{particl}&particl1, particl2&\textbf{whistle}&whistle1, whistle2,\\
        \textbf{cap}&cap1, cap2&&whistle3\\
        \textbf{clap}&clap1, clap2&\textbf{phone}&phone1, phone2,\\
        \textbf{claps}&claps1, claps2&&phone3, phone4\\
        \textbf{clip}&clip1, clip2&\textbf{toy}&toy1, toy2\\
        \textbf{bell} & bell1, bell2\\
        \wcline{1-4}
    \end{tabular}%
    }
\end{table}
\begin{table*}[t!]
\caption{SDRi [dB] for extracted signals}
\vspace{2pt}
\label{table:result}
\centering
\resizebox{\linewidth}{!}{%
\begin{tabular}{@{}llccccc@{}}
    \wcline{1-7}
     & \\[-8pt]
      & & \multicolumn{5}{c}{SNR} \\
     \cline{3-7}\\[-8pt]
     Dataset& Method&\SI{-10}{\dB} & \SI{-5}{\dB} & \SI{0}{\dB} & \SI{5}{\dB} & \SI{10}{\dB}\\
    \cline{1-7}
    & \\[-8pt]
    \multirow{3}{*}{Inter-superclass dataset} & Superclass-conditioned method&$5.11 \pm 3.02$ & $4.72 \pm 2.75$ & $4.06 \pm 2.55$ & $2.70 \pm 2.13$ & $1.33 \pm 2.12$ \\
     & Subclass-conditioned method&$5.06 \pm 2.97$ & $4.75 \pm 2.85$ & $4.04 \pm 2.52$ & $2.81 \pm 2.31$ &  $1.25 \pm 2.09$\\
     & {\bf Onomatopoeia-conditioned method} &$4.63 \pm 2.58$ & $4.57 \pm 2.69$ & $4.02 \pm 2.53$ & $2.77 \pm 2.22$ & $1.41 \pm 2.12$ \\
    \cline{1-7}
    & \\[-8pt]
    \multirow{3}{*}{Intra-superclass dataset} & Superclass-conditioned method&$2.05 \pm 2.37$ & $1.97 \pm 2.40$ & $1.86 \pm 2.38$ & $1.50 \pm 2.19$ & $0.82 \pm 1.89$ \\
     & Subclass-conditioned method&$5.03 \pm 2.56$ & $4.77 \pm 2.59$ & $4.19 \pm 2.45$ & $2.74 \pm 2.12$ &  $1.26 \pm 2.06$\\
     & {\bf Onomatopoeia-conditioned method} &$5.61 \pm 2.78$ & $5.36 \pm 2.75$ & $4.73 \pm 2.52$ & $3.10 \pm 2.27$ & $1.42 \pm 2.06$ \\
    \cline{1-7}
    & \\[-8pt]
    \multirow{3}{*}{Intra-subclass dataset} & Superclass-conditioned method&$2.03 \pm 2.40$ & $2.06 \pm 2.54$ & $1.87 \pm 2.37$ & $1.49 \pm 2.09$ & $0.79 \pm 1.98$ \\
     & Subclass-conditioned method&$3.14 \pm 2.78$ & $3.09 \pm 2.77$ & $2.84 \pm 2.63$ & $2.21 \pm 2.29$ &  $1.01 \pm 2.12$\\
     & {\bf Onomatopoeia-conditioned method} &$5.83 \pm 2.43$ & $5.68 \pm 2.53$ & $5.11 \pm 2.58$ & $3.34 \pm 2.24$ & $1.64 \pm 2.02$ \\
    \wcline{1-7}
\end{tabular}%
}
\end{table*}
\begin{figure*}[t!]
\vspace{5pt}
\centering
\includegraphics[scale=0.75]{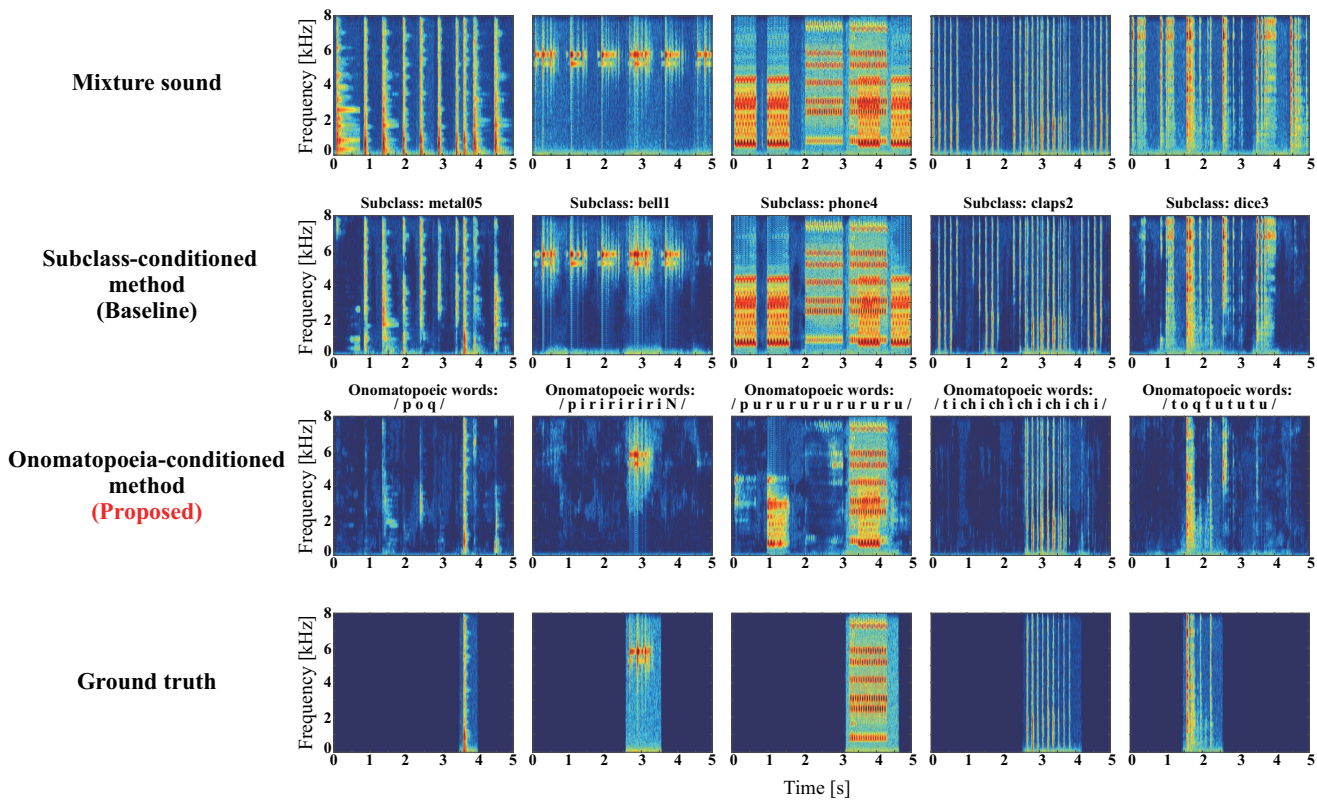}
\vspace{-11pt}
\caption{Examples of environmental sound extraction using intra-subclass dataset. Mixture spectrogram (first row), results of subclass-conditioned sound extraction (second row), results of onomatopoeia-conditioned sound extraction (proposed) (third row), and ground truth spectrogram (fourth row).}
\label{fig:spectro_result}
\end{figure*}

\subsection{Training and evaluation setup}
\label{sec:setup}
\vspace{-5pt}
Table~\ref{table:experiment} shows the experimental conditions and parameters used for the proposed method (onomatopoeia-conditioned method).
As baselines, we also evaluated the methods with which the target sound is conditioned by the superclass or subclass sound-event class.
We used the one-hot representation of the label for $\bmit{o}$ in (\ref{eq:word_embedding}) instead of the word embeddings.

To evaluate each method, we used signal-to-distortion ratio improvement (SDRi) \cite{SDR_TASLP_2006} as an evaluation metric.
SDRi is defined as the difference between the SDR of the target sound to the mixture and that of the target sound to the extracted sound as follows:
\begin{align}
    \mbox{SDRi} = 10\log_{10}\left(\frac{\norm{\bm{y}}^2}{\norm{\bm{y} - \hat{\bm{y}}}^2}\right) - 10\log_{10}\left(\frac{\norm{\bm{y}}^2}{\norm{\bm{y} - \bm{x}}^2}\right).
\end{align}
We conducted evaluations regarding SDRi on each of the three evaluation datasets introduced in Sec.~\ref{sec:dataset}.

\subsection{Experimental results}
\label{sec:results}
\vspace{-5pt}
Table~\ref{table:result} shows the SDRi on each evaluation dataset.
We observed that the superclass-conditioned method performed well on the inter-superclass dataset but performed poorly on the intra-superclass and intra-subclass datasets.
We also observed that the subclass-conditioned method performed well on the inter-superclass and intra-superclass datasets but did not on the intra-subclass dataset.
These results indicate that the performance of sound extraction using an event class as a condition is highly dependent on the fineness of the class definition.
The onomatopoeia-conditioned method showed almost the same SDRi on the three datasets.
This suggests that an onomatopoeic word can behave like a more fine-grained class than the subclasses, even though it does not require any special domain knowledge for labeling.

Figure \ref{fig:spectro_result} shows the spectrograms of the extracted sounds using the subclass-conditioned and onomatopoeia-conditioned methods.
For this visualization, we used five samples in the intra-subclass dataset with \SI{0}{\dB}.
We observed that the subclass-conditioned method left a significant amount of non-target sounds, while the onomatopoeia-conditioned method extracted only the target sound.
Although the onomatopoeia-conditioned method performed better than the superclass- and subclass-conditioned methods, it still did not perform well when the target sound was highly overlapped with interference sounds (cf. ``Subclass: Phone4'' in Fig.~\ref{fig:spectro_result} ).
As a result, the mixtures with high overlap ratios resulted in small SDRi and the mixtures with low overlap ratios resulted in large SDRi, and thus the standard deviations in Table \ref{table:result} are large overall.
The extraction of overlapping sounds requires further study.
The extracted sounds are available on our web page\footnote{\url{ https://y-okamoto1221.github.io/Sound_Extraction_Onomatopoeia/}}.

\section{CONCLUSION}
\label{sec:conclusion}
\vspace{-5pt}
We proposed an environmental-sound-extraction method using onomatopoeic words.
The proposed method estimates a time-frequency mask of the target sound specified by an onomatopoeic word with the U-Net encoder-decoder architecture.
The experimental results indicate that our proposed method extracts specific sounds from mixture sounds by using an onomatopoeic word as a condition.
Our proposed method outperformed conventional methods that use a sound-event class as a condition.
The results indicate that onomatopoeic words can behave like more fine-grained classes than sound-event classes, even though it does not require any special domain knowledge for labeling.
In the future, it will be necessary to verify the effectiveness of the proposed method for onomatopoeic words assigned by speakers of different languages.

\vfill\pagebreak


\bibliographystyle{IEEEtran}
\bibliography{refs}

\begin{thebibliography}{10}
\providecommand{\url}[1]{#1}
\csname url@samestyle\endcsname
\providecommand{\newblock}{\relax}
\providecommand{\bibinfo}[2]{#2}
\providecommand{\BIBentrySTDinterwordspacing}{\spaceskip=0pt\relax}
\providecommand{\BIBentryALTinterwordstretchfactor}{4}
\providecommand{\BIBentryALTinterwordspacing}{\spaceskip=\fontdimen2\font plus
\BIBentryALTinterwordstretchfactor\fontdimen3\font minus
  \fontdimen4\font\relax}
\providecommand{\BIBforeignlanguage}[2]{{%
\expandafter\ifx\csname l@#1\endcsname\relax
\typeout{** WARNING: IEEEtran.bst: No hyphenation pattern has been}%
\typeout{** loaded for the language `#1'. Using the pattern for}%
\typeout{** the default language instead.}%
\else
\language=\csname l@#1\endcsname
\fi
#2}}
\providecommand{\BIBdecl}{\relax}
\BIBdecl

\bibitem{Imoto_AST_2018}
K.~Imoto, ``Introduction to acoustic event and scene analysis,'' \emph{Acoust.
  Sci. Tech.}, vol.~39, no.~3, pp. 182--188, 2018.

\bibitem{zhang_CHIIR_2020}
Y.~Zhang, J.~Hu, Y.~Zhang, B.~Pardo, and Z.~Duan, ``Vroom!: A search engine for
  sounds by vocal imitation queries,'' in \emph{Proc. Conference on Human
  Information Interaction and Retrieval {\rm (}{CHIIR}{\rm )}}, 2020, pp.
  23--32.

\bibitem{Zhang_ICASSP_2016}
Y.~Zhang and Z.~Duan, ``{IMISOUND}: An unsupervised system for sound query by
  vocal imitation,'' in \emph{Proc. IEEE International Conference on Acoustics,
  Speech and Signal Processing {\rm (}{ICASSP}{\rm )}}, 2016, pp. 2269--2273.

\bibitem{Kim_ICASSP_2019}
B.~Kim and B.~Pardo, ``Improving content-based audio retrieval by vocal
  imitation feedback,'' in \emph{Proc. IEEE International Conference on
  Acoustics, Speech and Signal Processing {\rm (}{ICASSP}{\rm )}}, 2019, pp.
  4100--4104.

\bibitem{Zhang_TASLP_2019}
Y.~Zhang, B.~Pardo, and Z.~Duan, ``Siamese style convolutional neural networks
  for sound search by vocal imitation,'' \emph{IEEE/ACM Transactions on Audio,
  Speech, and Language Processing}, vol.~27, no.~2, pp. 429--441, 2019.

\bibitem{Ikawa_DCASE_2018}
S.~Ikawa and K.~Kashino, ``Acoustic event search with an onomatopoeic query:
  measuring distance between onomatopoeic words and sounds,'' in \emph{Proc.
  Workshop on Detection and Classification of Acoustic Scenes and Events {\rm
  (}{DCASE}{\rm )}}, 2018, pp. 59--63.

\bibitem{Lemaitre_JASA_2018}
G.~Lemaitre and D.~Rocchesso, ``On the effectiveness of vocal imitations and
  verbal descriptions of sounds,'' \emph{The Journal of the Acoustical Society
  of America}, vol. 135, no.~2, pp. 862--873, Feb. 2014.

\bibitem{Sundaram_AAAI_2006}
S.~Sundaram and S.~Narayanan, ``Vector-based representation and clustering of
  audio using onomatopoeia words,'' in \emph{Proc. American Association for
  Artificial Intelligence {\rm (}{AAAI}{\rm )} Symposium Series}, 2006, pp.
  55--58.

\bibitem{Ronneberger_MICCAI_2015}
O.~Ronneberger, P.~Fischer, and T.~Brox, ``U-{N}et: Convolutional networks for
  biomedical image segmentation,'' in \emph{Proc. International Conference on
  Medical Image Computing and Computer-Assisted Intervention {\rm
  (}{MICCAI}{\rm )}}, 2015, pp. 234--241.

\bibitem{Mesequer_ISMIR_2019}
G.~M.-Brocal and G.~Peeters, ``Conditioned-{U}-{N}et: Introducing a control
  control mechanism in the {U}-{N}et for multiple source separations,'' in
  \emph{Proc. International Society for Music Information Retrieval {\rm
  (}{ISMIR}{\rm )}}, 2019, pp. 159--165.

\bibitem{Sudo_IROS_2019}
Y.~Sudo, K.~Itoyama, K.~Nishida, and K.~Nakadai, ``Environmental sound
  segmentation utilizing {Mask} {U}-{N}et,'' in \emph{Proc. {IEEE}/{RSJ}
  International Conference on Intelligent Robots and Systems {\rm (}{IROS}{\rm
  )}}, 2019, pp. 5340--5345.

\bibitem{Jansson_ISMIR_2017}
A.~Jansson, E.~J. Humphrey, N.~Montecchio, R.~M. Bittner, A.~Kumar, and
  T.~Weyde, ``Singing voice separation with deep {U}-{N}et convolutional
  networks,'' in \emph{Proc. International Society for Music Information
  Retrieval {\rm (}{ISMIR}{\rm )}}, 2017, pp. 745--751.

\bibitem{Kong_ICASSP_2020}
Q.~Kong, Y.~Wang, X.~Song, Y.~Cao, W.~Wang, and M.~D. Plumbley, ``Source
  separation with weakly labelled data: an approach to computational auditory
  scene analysis,'' in \emph{Proc. IEEE International Conference on Acoustics,
  Speech and Signal Processing {\rm (}{ICASSP}{\rm )}}, 2020, pp. 101--105.

\bibitem{ochiai_INTERSPEECH_2020}
T.~Ochiai, M.~Delcroix, Y.~Koizumi, H.~Ito, K.~Kinoshita, and S.~Araki,
  ``Listen to what you want: Neural network-based universal sound selector,''
  in \emph{Proc. {\rm }{INTERSPEECH}{\rm }}, 2020, pp. 1441--1445.

\bibitem{Lee_ISMIR_2019}
J.~H. Lee, H.-S. Choi, and K.~Lee, ``Audio query-based music source
  separation,'' in \emph{Proc. International Society for Music Information
  Retrieval {\rm (}{ISMIR}{\rm )}}, 2019, pp. 878--885.

\bibitem{Kavalerov_WASPPA_2019}
I.~Kavalerov, S.~Wisdom, H.~Erdogan, B.~Patton, K.~Wilson, J.~L. Roux, and
  J.~R. Hershey, ``Universal sound separation,'' in \emph{Proc. IEEE Workshop
  on Applications of Signal Processing to Audio and Acoustics {\rm
  (}{WASPAA}{\rm )}}, 2019, pp. 175--179.

\bibitem{Slizovskaia_ICASSP_2109}
O.~Slizovskaia, L.~Kim, G.~Haro, and E.~Gomez, ``End-to-end sound source
  separation conditioned on instrument labels,'' in \emph{Proc. IEEE
  International Conference on Acoustics, Speech and Signal Processing {\rm
  (}{ICASSP}{\rm )}}, 2019, pp. 306--310.

\bibitem{Luo_TASLP_2019}
Y.~Luo and N.~Mesgarani, ``Conv-{T}as{N}et: Surpassing ideal time–frequency
  magnitude masking for speech separation,'' \emph{IEEE/ACM Transactions on
  Audio, Speech, and Language Processing}, vol.~27, no.~8, pp. 1256--1266,
  2019.

\bibitem{Smaradis_WASPAA_2019}
P.~Smaragdis and G.~J. Mysore, ``Separation by ``humming'': User-guided sound
  extraction from monophonic mixtures,'' in \emph{Proc. IEEE Workshop on
  Applications of Signal Processing to Audio and Acoustics {\rm (}{WASPAA}{\rm
  )}}, 2019, pp. 69--72.

\bibitem{Griffin_TASSP_1984}
D.~Griffin and J.~Lim, ``Signal estimation from modified short-time {F}ourier
  transform,'' \emph{{IEEE} Transactions on Acoustics, Speech, and Signal
  Processing}, vol.~32, no.~2, pp. 236--243, 1984.

\bibitem{Nakamura_AST_1999}
S.~Nakamura, K.~Hiyane, F.~Asano, and T.~Endo, ``Sound scene data collection in
  real acoustical environments,'' \emph{The Journal of the Acoustic Society of
  Japan (E)}, vol.~20, no.~3, pp. 225--231, 1999.

\bibitem{okamoto_DCASE_2020}
Y.~Okamoto, K.~Imoto, S.~Takamichi, R.~Yamanishi, T.~Fukumori, and
  Y.~Yamashita, ``{RWCP-SSD-Onomatopoeia}: Onomatopoeic words dataset for
  environmental sound synthesis,'' in \emph{Proc. Workshop on Detection and
  Classification of Acoustic Scenes and Events {\rm (}{DCASE}{\rm )}}, 2020,
  pp. 125--129.

\bibitem{RAdam_ICLR_2020}
L.~Liu, H.~Jiang, P.~He, W.~Chen, X.~Liu, J.~Gao, and J.~Han, ``On the variance
  of the adaptive learning rate and beyond,'' in \emph{Proc. International
  Conference on Learning Representation {\rm (}{ICLR}{\rm )}}, 2020, pp. 1--13.

\bibitem{SDR_TASLP_2006}
E.~Vincent, R.~Gribonval, and C.~F\'{e}votte, ``Performance measurement in
  blind audio source separation,'' \emph{IEEE/ACM Transactions on Audio,
  Speech, and Language Processing}, vol.~14, no.~4, pp. 1462--1469, 2006.

\end{thebibliography}

\end{document}